\documentclass[gmd, manuscript]{copernicus}

\usepackage{subfig}
\usepackage{rotating}

\usepackage{url} 
\usepackage{lineno}
\usepackage{soul}
\usepackage{amsmath}

\graphicspath{{images/}}

\begin{document}
\begin{nolinenumbers}

\title{Identifying Lightning Processes in ERA5 Soundings with Deep Learning}


\Author[1, 2][gregor.ehrensperger@uibk.ac.at]{Gregor}{Ehrensperger} 
\Author[2]{Thorsten}{Simon}
\Author[2]{Georg J.}{Mayr}
\Author[1]{Tobias}{Hell}

\affil[1]{Data Lab Hell GmbH, Austria}
\affil[2]{Department of Atmospheric and Cryospheric Sciences, University of Innsbruck}




\runningtitle{Identifying Lightning Processes in ERA5 Soundings with Deep Learning}
\runningauthor{G. Ehrensperger et al.}

\received{}
\pubdiscuss{} 
\revised{}
\accepted{}
\published{}

\firstpage{1}

\maketitle


\begin{abstract}
Atmospheric environments favorable for lightning and convection are commonly
represented by proxies or parameterizations based on expert knowledge such as
CAPE, wind shears, charge separation, or combinations thereof.
Recent developments in the field of machine learning, high resolution reanalyses,
and accurate lightning observations open possibilities for identifying tailored
proxies without prior expert knowledge.

To identify vertical profiles favorable for lightning,
a deep neural network links ERA5 vertical profiles of cloud physics, mass field variables and wind to lightning location data from the \textit{Austrian Lightning Detection \& Information System} (ALDIS), which has been transformed to a binary target variable labeling the ERA5 cells as \textit{cells with lightning activity} and \textit{cells without lightning activity}.
The ERA5 parameters are taken on model levels beyond the tropopause forming an input layer of approx.\ 670~features.
The data of 2010--2018 serve as training/validation.

On independent test data, 2019, the deep network outperforms a reference with features based on meteorological expertise.
SHAP values highlight the atmospheric processes learned by the network which identifies cloud ice and snow content in the upper and mid-troposphere as very relevant features.
As these patterns correspond to the separation of charge in thunderstorm cloud, the deep learning model can serve as physically meaningful description of lightning.

Depending on the region, the neural network also exploits the vertical wind or mass profiles to correctly classify cells with lightning activity.
\end{abstract}


\introduction
Lightning affects many fields of our everyday's life. Cloud-to-ground flashes
might hit infrastructure such as wind turbines \citep{becerra2018} and power
lines \citep{cummins1998} and thus cause power outages.
Humans might get injured \citep{ritenour2008} or even die \citep{holle2016} after being hit by lightning.
Wildfires \citep{reineking2010} release carbon dioxide into the climate system,
and thus limit the biosphere's capacity to store carbon dioxide.
Lightning also affects the climate system by producing nitrogen oxides which play a key role in ozone conversion and acid rain production \citep{decaria2005}.
Ozone is an important greenhouse gas and changes in concentration can lead to warming or cooling of the atmosphere. 
Thus, understanding of lightning is also an important factor in climate change research \citep{finney2018}.

Given lightning's impact and that an average of 46 flashes are occurring around the globe every second \citep{cecil2014} it is desirable to have models of the atmosphere
capable to simulate lightning and its underlying dynamic processes down to the
resolved scales of the numeric model.
Beyond the resolved scales one relies on so called proxies \textit{or} parameterization to further describe lightning.
The term \textit{proxy} is commonly used for quantities derived from atmospheric model output \textit{after} numeric computations.
\textit{Parameterizations} mean the description of lightning \textit{while} numeric computations of the atmosphere model.

Proxies are frequently applied to assess historic and future behavior
of convection and lightning.
Popular proxies are cloud top height \citep{price1992}, cloud ice flux \citep{finney2014}, CAPE times precipitation \citep{romps2018}, or the lightning potential index \citep{brisson2021}.
Though, these proxies perform reasonably good \citep{tippett2019}, there is a need for more complex or holistic proxies, as the behavior of lightning in a changing climate is still uncertain \citep{murray2018}.
Another application that makes clear that more research on the description of
lightning is needed in the field of operational weather forecasting.
Experience shows, for instance, that CAPE needs to be adapted to local conditions in order to perform well \citep{groenemeijer2019ecmwfmemo}.

Parameterizations are an internal part of numeric models, as they emulate sub-scale processes that cannot be resolved due the discretization of governing equations.
Therefore, the emulated processes give feedback to the other processes, also on larger scales, within the atmospheric model.
For instance, \citet{tost2007} showed that modeled nitrogen oxide is sensitive to lightning parameterizations in numerical models.
Next to the classic description of lightning using cloud top height \citep{price1992}, parameterizations have been developed using polynomial regression \citep{allen2002} and schemes based on hydrometeors in the mixed-phase region which is important for cloud-resolving models \citep{mccaul2009}.
A comparison of several parameterizations using a superparameterized model is given by \citet{charn2021}.
Recently, the ECMWF launched a product for total lightning densities expressed as a function of hydrometeors contents, CAPE, and (convective) cloud-base height output by the convective parameterization \citep{lopez2016}.

In recent years machine learning (ML) approaches have been proposed to describe convection and lightning.
Forty preselected single level parameters from ERA5 were processed by artificial neural networks and
gradient boosting machines for lightning in parts of Europe and Sri Lanka \citep{ukkonen2017, ukkonen2019}.
The authors also bring up the idea to feed ERA5's model level parameters directly to an appropriate ML tool, i.e.\ neural network.
Other studies tested random forests for regions such as the Hubei Province in China \citep{shi2022} or the Southern Great Plains \citep{shan2023} and generalized additive models (GAM) for the European Alps \citep{simon2023}.
All these studies confirm that the use of ML approaches for the description of lightning is promising. 
In concurrent research, also \textit{explainable artificial intelligence} (XAI) techniques are used to move towards understanding the underlying reasoning of complex AI models and shows encouraging results in Earth System Sciences applications \citep{barnes2020, dutta2022, hilburn2021, mayer2021, stirnberg2021, toms2021}. 
\citet{silva2022} use XGBoost classification trees to explore when the NASA Goddard Earth Observing System model of lightning flash occurrence shows weaknesses and apply \textit{Shapley additive explanations} (SHAP) to describe which meteorological drivers are related to the model errors.
They found that these errors  are strongly related to convection in the atmosphere and certain characteristics of the land surface.

This paper builds upon these studies and aims at finding a \emph{holistic} description of lightning.
Supervised deep learning harvests temporally and vertically well resolved ERA5 soundings of atmospheric dynamics and cloud physics to explain observations from the \textit{Austrian Lightning Information \& Detection System} (ALDIS).
The pattern found in ERA5 serve as proxy, but could also guide towards a parameterization of lightning.
Using ERA5 on model levels comes with the benefit that a complete picture of the atmosphere is considered to find patterns explaining lightning.
As the approach sees the \emph{raw model atmosphere}, no expert parameters diagnosed from the model levels are used as inputs, the study also answers
whether deep learning can identify physically meaningful patterns within the ERA5
sounding to describe lightning processes.

The region of interest are the eastern Alps which are characterized by complex terrain.
Atmospheric dynamics on a gamut of scales interacting with topography, which lead to various meso-scale \citep{feldmann2021} and local processes \citep{houze2012} that can trigger convection and lightning.
This study focuses on lightning during the peak phase of the warm season (June, July, August) which differs fundamentally in the underlying dynamic processes to lightning during the cold season \citep{morgenstern2022}.
\cite{morgenstern2023} show that there are different environments, either dominated by wind-field or mass-field variables, that favor lightning depending on the region. 

The paper is structured as follows.
Section~\ref{sec:data} presents both the lightning detection data and the atmospheric reanalyses, which both enter the supervised deep learning approach and a reference model (Sect.~\ref{sec:methods}).
Additionally, Section~\ref{sec:methods} illustrates the methods to analyze performance and explainability of the deep learning approach.
The results of these analyses are given in Section~\ref{sec:results}.
Section~\ref{sec:discussion} discusses the physical patterns identified by the method, scrutinizes further applications and research that is made possible with the novel insights and finally concludes the study.

\section{Data}\label{sec:data}

Two data sets build the foundation for this supervised machine learning task.
First, the observational data from the lightning location system ALDIS
(Sect.~\ref{sec:data:aldis}) is used to derive the labels distinguishing
cells with and without lightning activity.
Second, pseudo soundings from ERA5 (Sect.~\ref{sec:data:era5}) serve as
input for the deep learning approach.
Spatially, the grid centers range from $8.25^{\circ}E$ to $16.75^{\circ}E$
and from $45.25^{\circ}N$ to $49.75^{\circ}N$.
Temporally, data for the meteorological summers (June, July, August) from 2010 to 2019 are available.

\subsection{Lightning Detection Data}\label{sec:data:aldis}

The Austrian Lightning Detection \& Information System (ALDIS) is part of the European effort EUCLID \citep{schulz2016}.
Cloud-to-ground flashes with a current of $>\,15\,\operatorname{kA}$ or $<\,-2\,\operatorname{kA}$ are aggregated to the
spatio-temporal grid cells of ERA5 (Sect.~\ref{sec:data:era5}).
Each cell has a horizontal extent of approx.\ $30\,\operatorname{km}\,\times\,30\,\operatorname{km}$ and temporally of one hour.
If at least one flash has been detected in such a grid cell, then the cell is labeled as \textit{cell with lightning activity}.
Otherwise, if not a single flash has been detected, the cell is labeled as \textit{cell without lightning activity}.

\subsection{Atmospheric Reanalysis}\label{sec:data:era5}

ECMWF's fifth reanalyses, ERA5 \citep{hersbach2020}, is available at a horizontal resolution of $0.25^{\circ}$ and temporally of 1 hour.
Vertically it consists of 137 hybrid model levels that align with topography
near ground and approach isobars in the upper
atmosphere (see \url{https://confluence.ecmwf.int/display/UDOC/L137+model+level+definitions}).
On these model levels nine parameters (Tab.~\ref{tab:era5}) are available to describe the state of the atmosphere.
Next to classical parameters, temperature, specific humidity and three-dimensional winds, ERA5 provides a description of liquid and solid water particles in clouds, i.e.\ the specific content of ice, snow (including graupel), liquid water, and rain.
For this study, these parameters are used on the lowest 74 model levels, spanning from level number 64 (approx.~$15\,000\,\operatorname{m}$ geopotential height) to level number 137 ($10\,\operatorname{m}$ above ground).

\begin{table}
\caption{ERA5 parameters on model levels.}
\centering
\begin{tabular}{l c c c}
\hline
Name & Short Name & Units & Parameter ID  \\
\hline
  Temperature & $\mathtt{t}$ & $K$ & 130   \\
  Specific humidity & $\mathtt{q}$ & $kg\,kg^{-1}$ & 133   \\
  U component of wind & $\mathtt{u}$ & $m\,s^{-1}$ & 131   \\
  V component of wind & $\mathtt{v}$ & $m\,s^{-1}$ & 132   \\
  Vertical velocity & $\mathtt{w}$ & $Pa\,s^{-1}$ & 135   \\
  Specific rain water content & $\mathtt{crwc}$ & $kg\,kg^{-1}$ & 75   \\
  Specific snow water content & $\mathtt{cswc}$ & $kg\,kg^{-1}$ & 76   \\
  Specific cloud liquid water content & $\mathtt{clwc}$ & $kg\,kg^{-1}$ & 246   \\
  Specific cloud ice water content & $\mathtt{ciwc}$ & $kg\,kg^{-1}$ & 247   \\
\hline
\end{tabular}
\label{tab:era5}
\end{table}

\subsection{Composition of Datasets}

The two data sets are merged in order to obtain a tabular data shape.
Each row of this tabular data refers to a spatio-temporal grid cell.
Thus, it can be indexed by the longitude and latitude of its center as
well as its hourly time stamp. Each row is either labeled as cell with lightning activity or without lightning activity.
The nine ERA5 parameters (Tab.~\ref{tab:era5}) on their 74 model levels enter the tabular data such that each resulting column refers to an \textit{individual}
parameter on an \textit{individual} level, making up a total of $9\times74=666$
ERA5 feature columns.
Further, each row is complemented with the information of the \textit{hour of the day} and \textit{day of the season} to account for diurnal and seasonal variations, respectively.
Finally, the model topography is added as another column.

\section{Methods}\label{sec:methods}

To avoid incorporating expert knowledge by using specialized deep learning architectures, a classical fully connected deep neural network (Sect.~\ref{sec:methods:dla}) is used to fit a model which capable of distinguishing whether a specific spatio-temporal grid cell corresponds to a lightning cell.
To make sure that the neural network can model lightning sufficiently well, the resulting outputs are compared to those of a state-of-the-art reference model (Sect.~\ref{sec:methods:rm}) on unseen test data.
Finally, insights into the patterns exploited by the trained model are gained by applying Shapley additive explanations (Sect.~\ref{sec:methods:expl}). 

\subsection{Deep Learning Approach}\label{sec:methods:dla}
A rather general fully connected neural network was designed, consisting of eight hidden layers with $512 \times 512 \times 512 \times 512 \times 128 \times 128 \times 128 \times 16$ nodes.
Leaky rectified linear unit (leaky ReLU) is used as activation function for all hidden layers.
The input dimension is predetermined by the number of input features and thus equals $671$ (nine atmospheric variables on 74 levels, longitude, latitude, hour of the day, day of the season, and topography).
The dimension of the output layer equals one, as it solely classifies whether the cell is with or without lightning activity.
The model output is activated with the sigmoid function.
The input features are scaled, such that the nine atmospheric variables are standardized by considering the 74 levels altogether, prior training.
To prevent the model from overfitting, dropout \citep{srivastava2014} with a value of $0.15$ and early stopping with a patience of ten epochs are applied.
Binary cross entropy serves as loss function with a weight of approximately $41$ for positive events (flash occurrences) to address for the highly imbalanced data set.

\subsection{Reference Model}\label{sec:methods:rm}
For reference a generalized additive model \citep[GAM,][]{wood2017generalized} is used.
This model is trained on longitude, latitude, hour of the day, day of the season, topography and the atmospheric variables listed in Tab.~\ref{tab:refmodelvars}, which were derived from ERA5 soundings on meteorological expertise \citep{simon2023}.

\begin{table}
\caption{The reference model is trained using the following ten atmospheric variables.}
\centering
\begin{tabular}{l c c c}
\hline
Description & Short Name \\
\hline
Convective available potential energy & $\mathtt{cape}$  \\
Binary indicator whether cloud is present & $\mathtt{cloud\_exists}$  \\
Convective precipitation & $\mathtt{cp}$  \\
Mass of specific snow water content between the $-20^{\circ}C$ and $-40^{\circ}C$ isotherms & $\mathtt{cswc2040}$  \\
Cloud top height in height above ground & $\mathtt{cth}$  \\
Instantaneous surface sensible heat flux & $\mathtt{ishf}$  \\
Medium cloud cover & $\mathtt{mcc}$  \\
Total column supercooled liquid & $\mathtt{tcslw}$  \\
Mass of water vapor between the $-10^{\circ}C$ and $-20^{\circ}C$ isotherms & $\mathtt{wvc1020}$  \\
Two meter temperature & $\mathtt{2t}$  \\
\hline
\end{tabular}
\label{tab:refmodelvars}
\end{table}

Thus, the input dimension for the reference model is only $15$. The GAM is fitted using an algorithm tailored for gigadata \citep{wood2017gigadata}.

\subsection{Explainability}\label{sec:methods:expl}
While generalized additive models are interpretable by users \citep{lou2017}, interpretability research of deep neural networks still suffers many gaps \citep{zhang2021}. 
Deep SHAP \citep{lundberg2017} is utilized to gain insights into the patterns exploited by the neural network from Section~\ref{sec:discussion} and to understand the features contributing to the classification of a spatio-temporal cell as one exhibiting lightning activity.
SHAP is a game theoretic approach to explain the relation of input and output of any machine learning model.
It uses the concept of Shapley values \citep{shapley1952} to provide local interpretability by computing feature attributions which lead to the model's output for a given input.
Unfortunately, the computation time for calculating the exact Shapley values grows exponentially with the number of input features.
Common implementations for computing Shapley values use simplifications to ensure computational feasibility.

This work applies Deep SHAP which is a model agnostic method that leverages extra knowledge about the nature of deep neural networks to approximate Shapley values more efficiently but also assumes independence among the input features.
As in various applications, this property obviously is not fulfilled in the given data set but using a more accurate approximation as described in \citet{aas2021} is not feasible with the large number of input features.
Despite this formal prerequisite, SHAP values are successfully utilized in a variety of imaging tasks (e.g. in the medical field \citep{vandervelden2022}) although the independence assumption is also heavily violated when images are used as input and every pixel serves as an individual input feature.
Understanding the raw ERA5 vertical profiles as a collection of 1D-images it is safe to assume that Deep SHAP leads to sufficiently good approximations to the precise Shapley values and can therefore be applied to gain insights into the neural network trained in this study.

\section{Results}\label{sec:results}
This section first investigates the performance of the deep learning approach by comparing its output on unseen test data against observations and the output of the reference model (Sect.~\ref{sec:results:comparison}).
Next, the application of SHAP allows to gain insights about the vertical profiles exploited by the neural network which indicate the occurrence of lightning (Sect.~\ref{sec:results:expl}).

\subsection{Performance of the deep learning approach}\label{sec:results:comparison}
The neural network is trained as described in Sect.~\ref{sec:methods:dla} to distinguish whether a given spatio-temporal cell is a cell with or without lightning activity. 
To map the output of the model to a binary category, a threshold has to be defined.
The threshold is determined by maximizing the $\operatorname{F}_1$ score, which symmetrically represents precision and recall in one metric, on the validation set.

The reference model is fitted as described in Sect.~\ref{sec:methods:rm} and the threshold is computed following the same procedure. 

The resulting confusion matrices are displayed in Tab.~\ref{tab:conf_mat}. 
The neural network slightly outperforms the reference model in every category of the confusion matrix on previously unseen test data (year~2019).
This can also be seen by comparing the \textit{Matthew correlation coefficients} (mcc) of the two models, where $+1$ represents a perfect match between model output and observations, and $0$ no better than random guessing.
The deep learning model has an mcc of approximately $0.278$ and the reference model $0.237$. 

Previous studies have shown that descriptions of lightning based on ML models reproduce the observed diurnal cycle more realistically than simple proxies such as
CAPE \citep[Fig~2. in][]{simon2023}.
To investigate the ability of the introduced model to reproduce the diurnal cycle of lightning, the mean of the binary model output and the observations of the test year 2019 for each hour of the day is calculated for four different small subdomains (Fig.\ref{fig:cycles}).
In this case, the model's threshold is calibrated to align the average predicted and observed lightning frequencies of the validation set.  

The comparison reveals a good match of the shapes of the modeled and observed diurnal cycles.
In particular, the transition from the low values in the morning to the peak in the afternoons are well reproduced.
For three of the subdomains (Flatlands, Northern Alpine Rim and Southern Alpine Rim), the model slightly overestimates the observed occurrence probabilities, which is a result of different mean occurrences of lightning in the validation (used for finding the threshold) and test data (plotted).
The only larger deviation can be found during the late afternoon in the High Alps, where the model overestimates the observed diurnal cycle.
It should be noted that the curves of the reference model are not directly comparable to those presented in \citep[Fig. 2 in][]{simon2023}, despite using the same model architecture and input variables.
This discrepancy arises from the fundamentally different data strategies employed.
\cite{simon2023} implemented a cross-validation method across the entire dataset spanning 2010 to 2019, averaged the calibrated probabilities for lightning occurrences and evaluated the mean of the modeled diurnal cycles over the ten years of available data.
In contrast, the current study exclusively utilizes data from the years 2010 to 2018 to fit the model and threshold.
The model's performance and resulting diurnal cycles are then assessed using data exclusively from the full year 2019, which has previously not been seen.

\begin{table}
  \caption{Confusion matrices of the neural network model (left) and the reference model (right) on test year 2019.}
  \centering
{\small
  \begin{tabular}{cc|c|c|c|}
& \multicolumn{2}{c}{}& \multicolumn{2}{c}{\textbf{observed}}\\
& \multicolumn{1}{c}{}& \multicolumn{1}{c}{\textbf{yes}}&\multicolumn{1}{c}{\textbf{no}}\\
\cline{3-4}
\multicolumn{1}{c}{\multirow{2}{*}{\textbf{modeled}}}
&\textbf{yes}    & 14\,370 & 61\,446 \\
\cline{3-4}
&\textbf{no} & 15\,768 & 1\,374\,741 \\
\cline{3-4}
\end{tabular}
}
\quad
{\small
  \begin{tabular}{cc|c|c|c|}
& \multicolumn{2}{c}{}& \multicolumn{2}{c}{\textbf{observed}}\\
& \multicolumn{1}{c}{}& \multicolumn{1}{c}{\textbf{yes}}&\multicolumn{1}{c}{\textbf{no}}\\
\cline{3-4}
\multicolumn{1}{c}{\multirow{2}{*}{}}
&\textbf{yes}    & 12\,654 & 65\,176 \\
\cline{3-4}
&\textbf{no} & 17\,484 & 1\,371\,011 \\
\cline{3-4}
\end{tabular}
}
 \label{tab:conf_mat}
\end{table}

\begin{figure}
	\centering
	\includegraphics[width=0.9\textwidth, keepaspectratio]{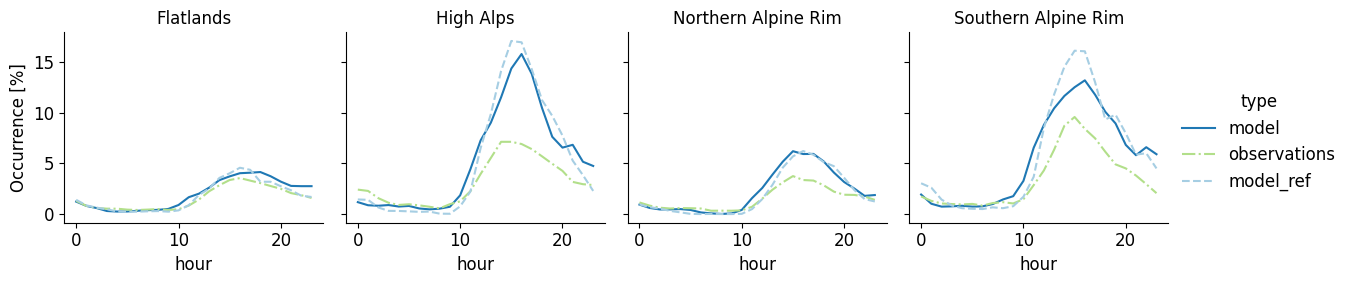}
	\caption{Diurnal cycles of the probability of a lightning occurring in a spatio-temporal cell.
	The binary model outputs of the neural network and reference model, as well as the binary observations have been averaged over four subdomains.}
	\label{fig:cycles}
\end{figure}

\subsection{Identifying patterns exploited by the deep learning model}\label{sec:results:expl}
The good performance of the deep learning approach motivates a closer look at what patterns the model has learned in order to distinguish between cells with and without lightning activity.
SHAP values (Sect.~\ref{sec:methods:expl}) indicate which inputs the neural network is particularly interested in. 
Since the goal is to find patterns which are valid throughout the full region used for training, and unbiased by the frequency of lightning within a specific spatial cell, the SHAP values are computed separately for each spatial cell. \footnote{In particular, within SHAP's \texttt{DeepExplainer} the full number of samples without lightning activity of the corresponding spatial cell are used as background data.}
Given a specific input, the SHAP values of all input features always sum up, with only small approximation errors, to the difference between the base value (derived from the expected model output based on background data) and the model output.
To better understand the underlying patterns, the SHAP values are scaled by dividing them by the difference between the base value of the corresponding spatial cell and the threshold at which a cell is classified as having lightning activity.
This implies that the model classifies a cell as having lightning activity as soon as the scaled SHAP values sum up to one or more, regardless of the underlying base value.
All plots in this paper illustrate these scaled SHAP values.
The aggregated results of the scaled SHAP values of correctly classified cells with lightning activity are visualized in Fig.~\ref{fig:shap}.

On average cloud ice (ciwc) and snow water content (cswc) contribute the most to the network's output.
Also note that ciwc with its lighter-weighted ice crystals is particularly interesting at a geopotential height of approx.\ $8000$ to $12000\,\operatorname{m}$ and cswc with its solid precipitation at approx.\ $3000$ to $10000\,\operatorname{m}$.

\begin{figure}
	\raggedright
	\includegraphics[trim=0 0 2500 0, clip, height=3.6cm]{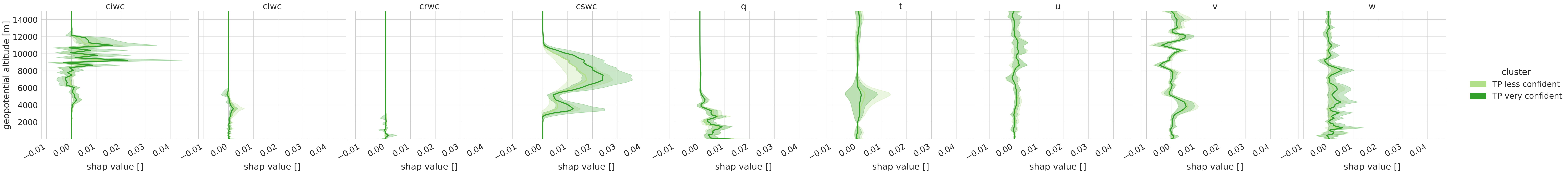}
	\includegraphics[trim=4000 0 0 0, clip, height=3.6cm]{profiles_agg_shap_q50_geopot_cl0-1_experimental_geopotheight}
    \vskip\baselineskip  
	\includegraphics[trim=0 0 4218 0, clip, height=3.6cm]{profiles_agg_shap_q50_geopot_cl0-1_experimental_geopotheight}%
	\includegraphics[trim=1850 0 350 0, clip, height=3.6cm]{profiles_agg_shap_q50_geopot_cl0-1_experimental_geopotheight}
	\caption{Scaled SHAP values for several variables (names on top of each subfigure) on correctly modeled lightning events (true positives).
	The two	colors represent the confidence (stratified by median) of the network in its output.
	The dark green color summarizes	the events where the network is very confident that a lightning event occurred.
	The light green	color summarizes the events where the network still modeled correctly, but with less confidence.
	The solid lines show the median of all observations and the colored areas highlight the 50\% quantiles.}
	\label{fig:shap}
\end{figure}

Taking a closer look (Fig.~\ref{fig:cloud}) at the ciwc and cswc at these altitudes, it is noticeable that the model exhibits greater confidence when ciwc and cswc values are substantially elevated. Furthermore, there is a tendency for the model to produce false positives during periods of high ciwc and cswc, while false negatives are more prevalent when these values are low compared to correctly classified lightning events.

\begin{figure}
  \centering
  \subfloat{
  	\subfloat{
    		\includegraphics[trim=0 0 280 0, clip, height=3.6cm]{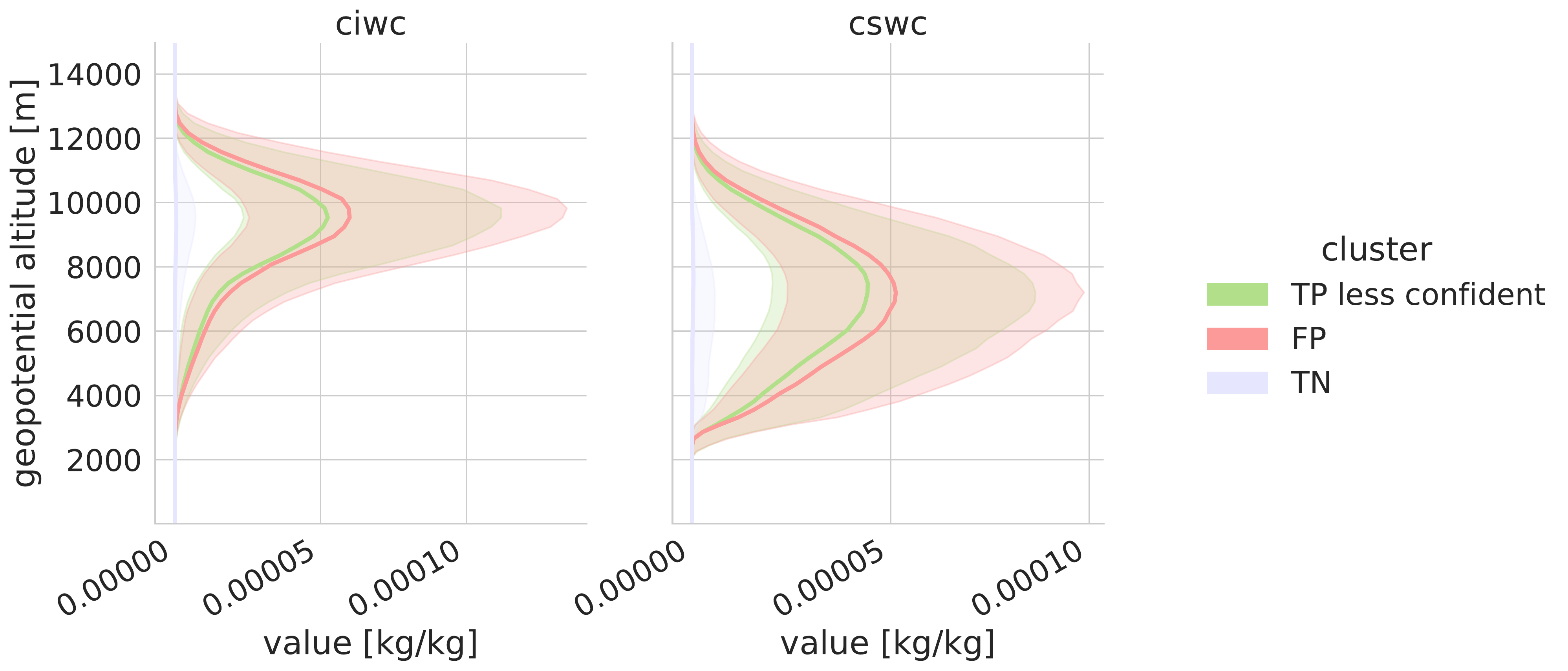}
    	}
  	\qquad
    \subfloat{
        \includegraphics[trim=120 0 0 0, clip, height=3.6cm]{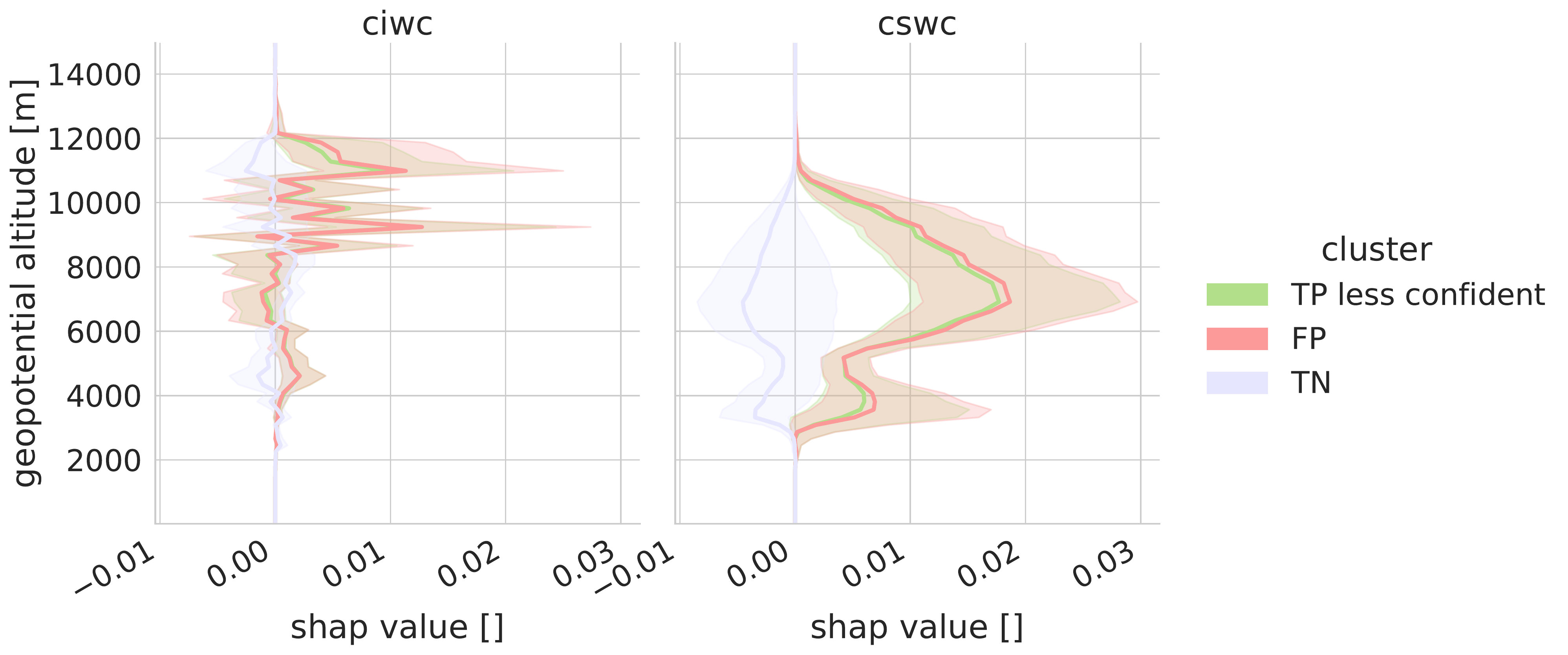}
    }
  }
  \vskip\baselineskip  
  \subfloat{
  	\subfloat{
    		\includegraphics[trim=0 0 280 0, clip, height=3.6cm]{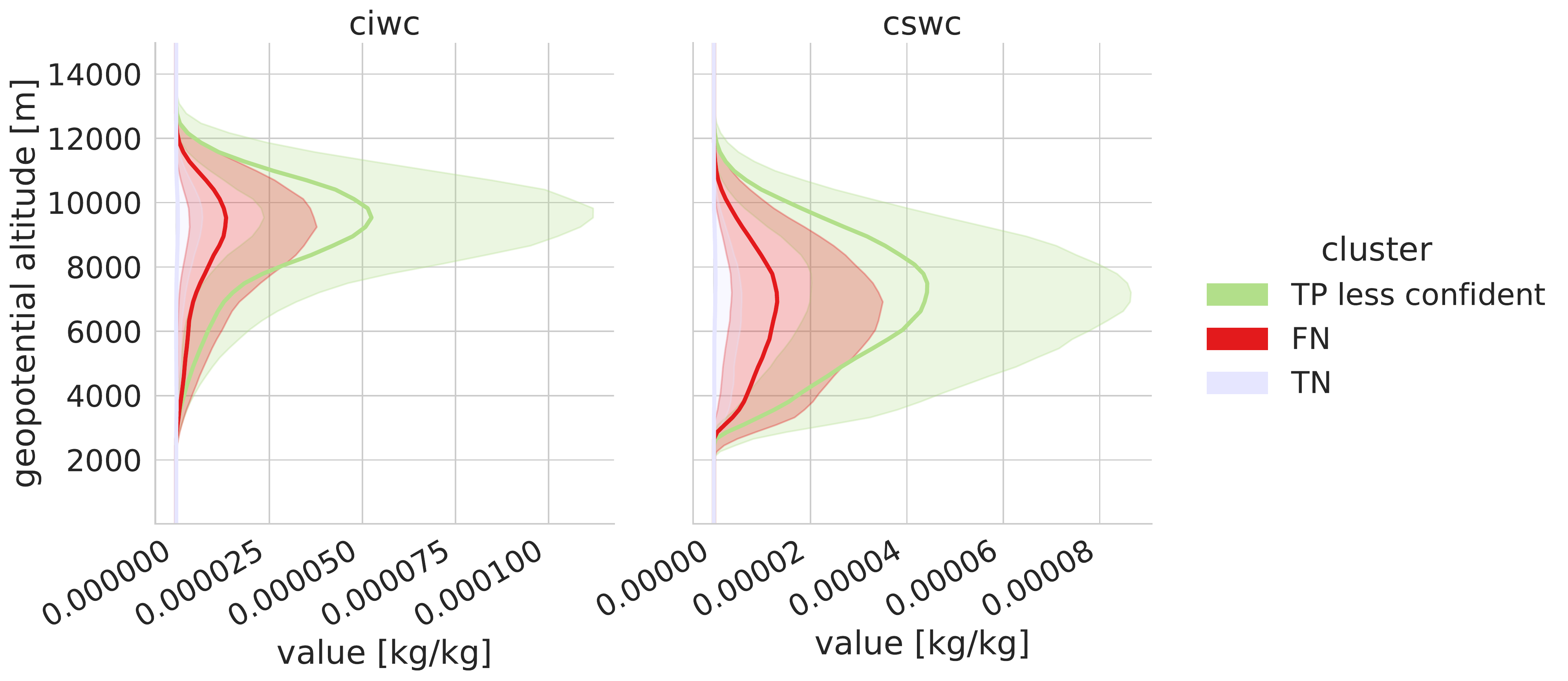}
  	}
  	\qquad
  	\subfloat{
	    \includegraphics[trim=120 0 0 0, clip, height=3.6cm]{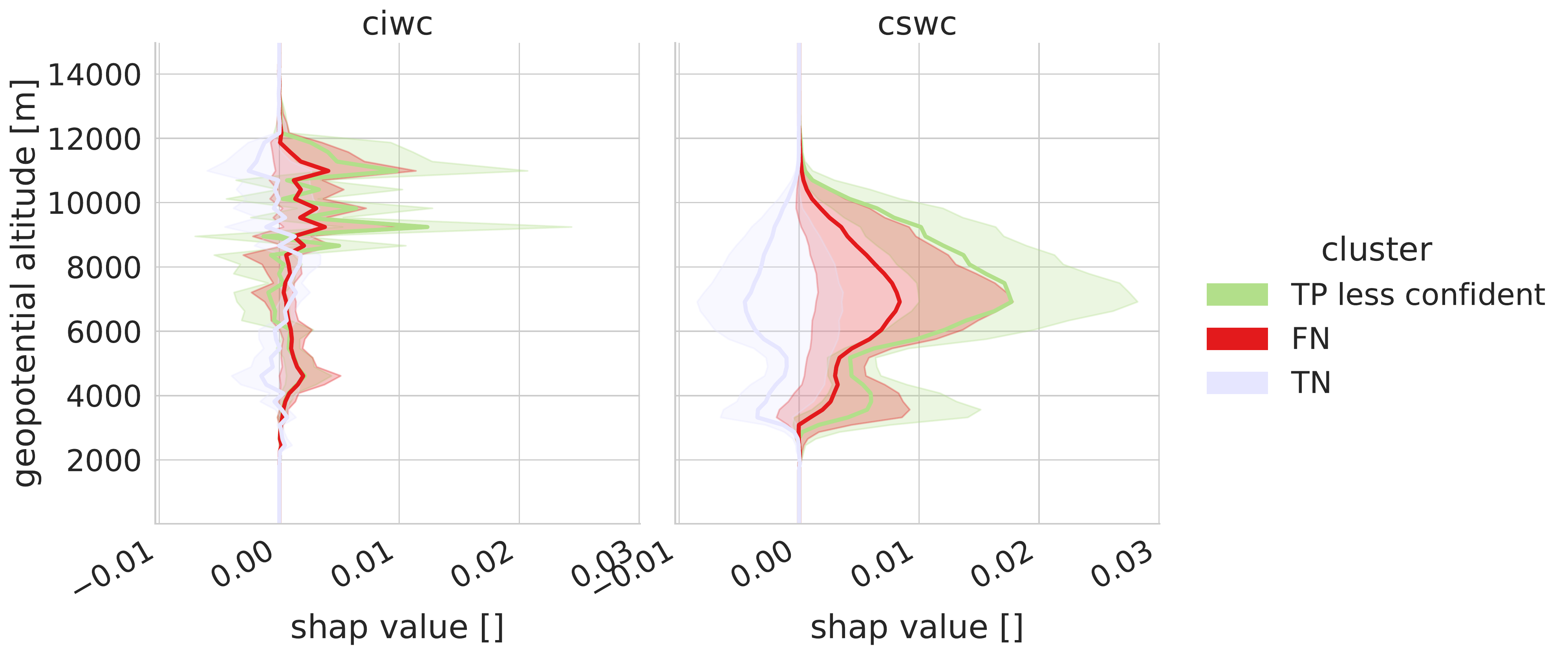}
  	}
  }
  
  \caption{The two left columns display the vertical profiles of the real feature labels, while the two right columns present the vertical profiles of the scaled SHAP values.
  The upper row illustrates less confident true positives (TP) compared to false positives (FP), while the lower row illustrates less confident true positives compared to false negatives (FN).
  True negatives (TN) are also included for reference.}
  \label{fig:cloud}
\end{figure}

While classifications where a cloudy atmosphere is the most dominantly exploited feature by the neural networks are the majority, grouping the results into three categories, following \cite{morgenstern2023}, reveals additional patterns:
\begin{itemize}
\item[\texttt{cloud}:] True positives where the sum of scaled SHAP values of ciwc, clwc, crwc and cswc exceeds $0.5$.
\item[\texttt{mass}:] True positives where the sum of scaled SHAP values of q and t exceeds $0.5$.
\item[\texttt{wind}:] True positives where the sum of scaled SHAP values of u, v and w exceeds $0.5$. 
\end{itemize}

Visualizing the vertical profiles of the scaled SHAP values (Fig.~\ref{fig:values:shap_grouped}) and the real feature values (Fig.~\ref{fig:values:features_grouped}) of these three groups it becomes clear that the mass-field lightning is characterized by warmer temperatures in the troposphere, a less stable stratification and copious amounts of water vapor in the lower troposphere.
Larger amounts of latent heat released by condensation as indicated by large clwc values in the lower troposphere combined with a weaker stratification result in more CAPE that can be released, which carries solid particles (ciwc) higher into the troposphere.
Heavier solid hydrometeors (cswc) peak further below.

Ice crystals and solid hydrometeors in wind-field lightning, on the other hand, are not transported that far up into the troposphere and they both peak in a similar altitude range.
The large-scale vertical velocity in the lower troposphere is high as is the horizontal wind speed -- particularly its southern component.
Temperatures and consequently specific humidity q are lower and the stratification is stabler than for the mass-field lightning.
All of this indicates forced lifting along (cold) fronts and topography.
Cold fronts in this region typically occur in southwesterly flow downstream of the trough axis, which explains the exceptional values of the v-component of the wind.
Charge separation consequently occurs on a tilted instead of a nearly vertical path as in mass field lightning, having earned this type of lightning the name \textit{tilted thunderstorm} \citep{brook1982, takeuti1978, takashi2019, wang2021}.

\begin{figure}
	\subfloat[scaled SHAP values]{
	\raggedright
	\begin{tabular}[b]{l}%
	\includegraphics[trim=0 0 2530 0, clip, height=3.6cm]{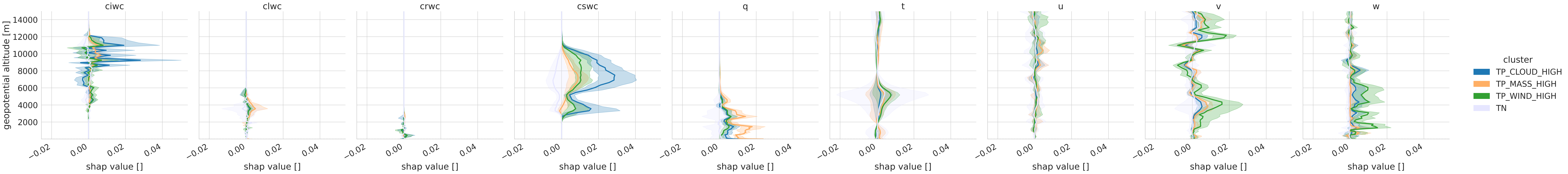}
	\includegraphics[trim=4000 0 0 0, clip, height=3.6cm]{profiles_agg_shap_q50_geopot_cl1-2-3-4_experimental_geopotheight.png} \\
	\includegraphics[trim=0 0 4212 0, clip, height=3.6cm]{profiles_agg_shap_q50_geopot_cl1-2-3-4_experimental_geopotheight.png}%
	\includegraphics[trim=1850 0 345 0, clip, height=3.6cm]{profiles_agg_shap_q50_geopot_cl1-2-3-4_experimental_geopotheight.png}
	\end{tabular}
	\label{fig:values:shap_grouped}
	}
	
	\medskip

	\subfloat[real features]{
	\raggedright
	\begin{tabular}[b]{l}%
	\includegraphics[trim=0 0 2530 0, clip, height=3.6cm]{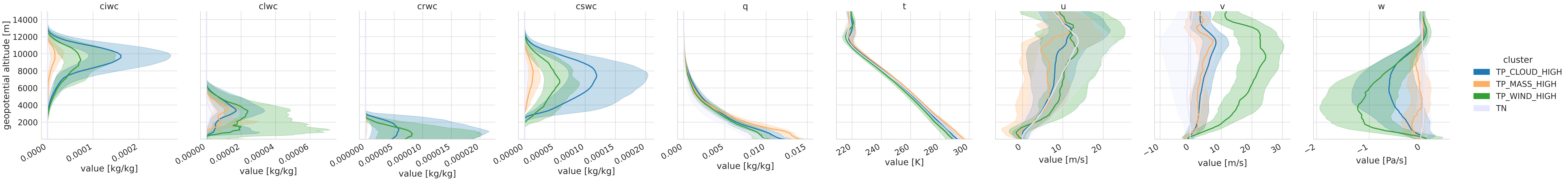}
	\includegraphics[trim=4000 0 0 0, clip, height=3.6cm]{profiles_agg_feature_q50_geopot_cl1-2-3-4_experimental_geopotheight.png} \\
	\includegraphics[trim=0 0 4215 0, clip, height=3.6cm]{profiles_agg_feature_q50_geopot_cl1-2-3-4_experimental_geopotheight.png}%
	\includegraphics[trim=1850 0 345 0, clip, height=3.6cm]{profiles_agg_feature_q50_geopot_cl1-2-3-4_experimental_geopotheight.png}
	\end{tabular}
	\label{fig:values:features_grouped}
	}
	\caption{Vertical profiles of the scaled SHAP-values (a) and real features (b) per variable with colors indicating true negatives and different groups of true positives (cloud-, mass-, wind-dominant).}
\end{figure}

Following the approach of \cite{morgenstern2023}, Fig.~\ref{fig:values:cloud} subdivides the cloud-dominant group into two subcategories: cloud-mass and cloud-wind.
This grouping is based on whether the aggregate of scaled SHAP values is greater for mass-related or wind-related parameters.

Lightning with large scaled SHAP and real values of the cloud variables seems to occur for both mass-field and wind-field lightning confirming the results from \cite{morgenstern2022, morgenstern2023}, who had used principal component analysis and clustering for identifying these categories.

\begin{figure}
	\subfloat[scaled SHAP values]{
	\raggedright
	\begin{tabular}[b]{l}%
	\includegraphics[trim=0 0 2580 0, clip, height=3.6cm]{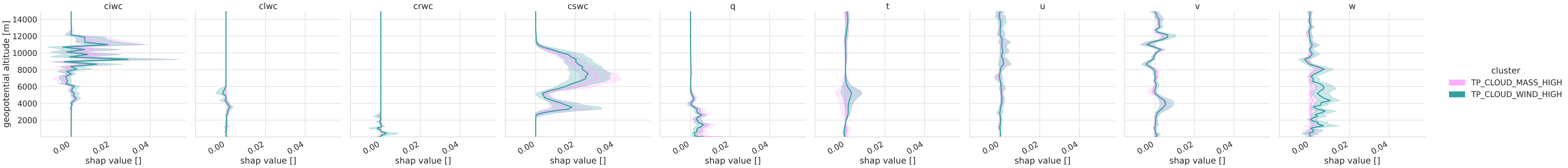}
	\includegraphics[trim=4000 0 0 0, clip, height=3.6cm]{profiles_agg_shap_q50_geopot_cl1-2_experimental_geopotheight.png} \\
	\includegraphics[trim=0 0 4280 0, clip, height=3.6cm]{profiles_agg_shap_q50_geopot_cl1-2_experimental_geopotheight.png}%
	\includegraphics[trim=1879 0 345 0, clip, height=3.6cm]{profiles_agg_shap_q50_geopot_cl1-2_experimental_geopotheight.png}
	\end{tabular}
	}
	\medskip

	\subfloat[real features]{
	\raggedright
	\begin{tabular}[b]{l}%
	\includegraphics[trim=0 0 2580 0, clip, height=3.6cm]{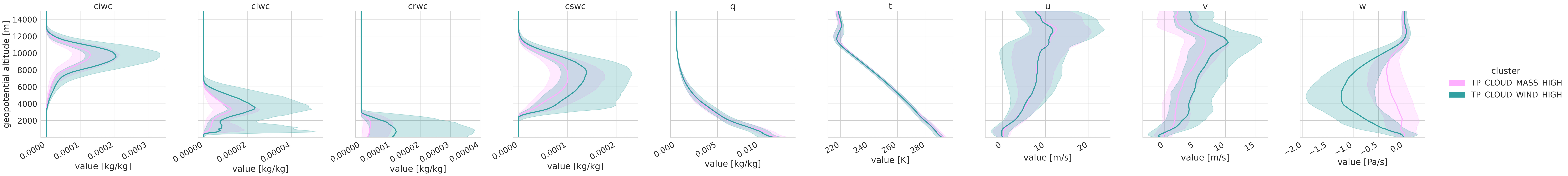}
	\includegraphics[trim=4000 0 0 0, clip, height=3.6cm]{profiles_agg_feature_q50_geopot_cl1-2_experimental_geopotheight.png} \\
	\includegraphics[trim=0 0 4280 0, clip, height=3.6cm]{profiles_agg_feature_q50_geopot_cl1-2_experimental_geopotheight.png}%
	\includegraphics[trim=1879 0 345 0, clip, height=3.6cm]{profiles_agg_feature_q50_geopot_cl1-2_experimental_geopotheight.png}
	\end{tabular}
	}
	\caption{Vertical profiles of the scaled SHAP-values (a) and real features (b) per variable with colors indicating cloud-mass and cloud-wind dominant true positives.}
	\label{fig:values:cloud}
\end{figure}

Fig.~\ref{fig:map:grouped} highlights the geographical regions, where cloud-mass-, cloud-wind-, mass-, or wind-dominant cells exhibiting lightning activity were classified.
Cloud-dominant cells with lightning activity are distributed across the entire map, but are particularly abundant along the primary chain of the Alps.
Mass-dominant cells are predominantly situated in Northern Italy and Slovenia.
Wind-dominant cells are primarily concentrated in the northwestern region of the Italian flat terrain, the Po Plain.

\begin{figure}
	\centering	

    \subfloat[cloud-mass-dominant]{
      \includegraphics[width=0.4\linewidth]{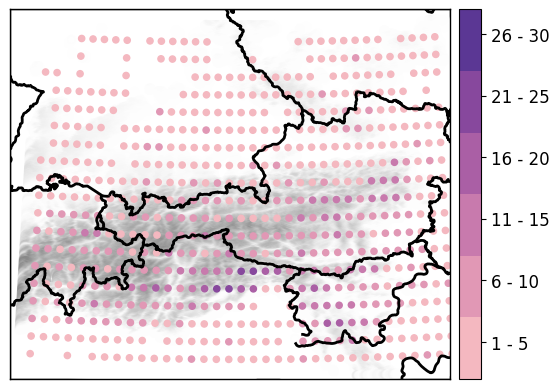}
    }
    \hfill
    \subfloat[cloud-wind-dominant]{
      \includegraphics[width=0.4\linewidth]{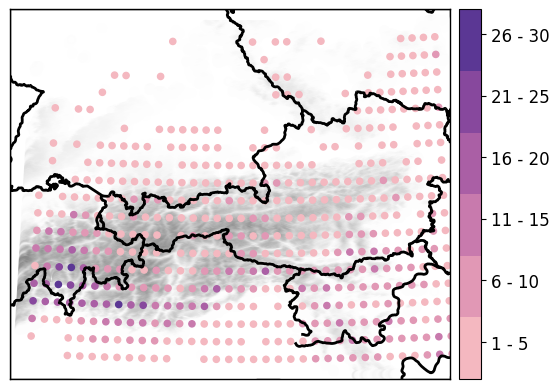}
    }
    
    \medskip

    \subfloat[mass-dominant]{
      \includegraphics[width=0.4\linewidth]{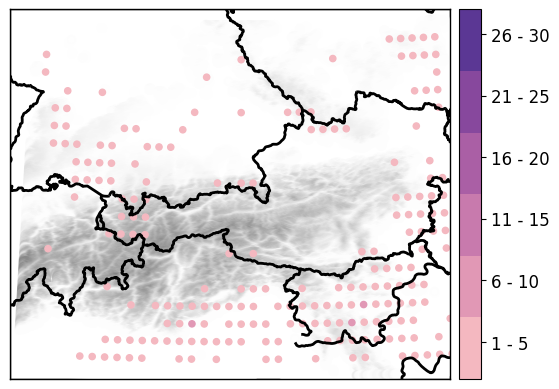}
    }
    \hfill
    \subfloat[wind-dominant]{
      \includegraphics[width=0.4\linewidth]{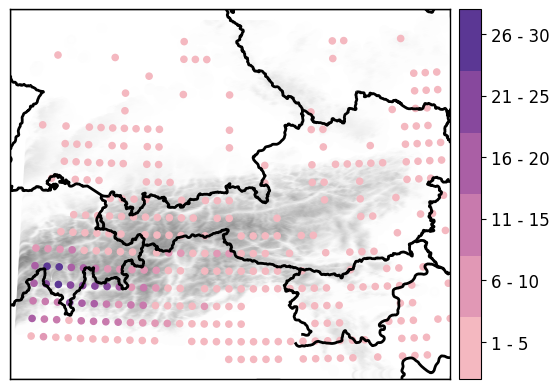}
    }
  
	\caption{The count of true positive classifications stratified by the variable group that is most dominant for each geographical location.
The data for the displayed topography layer is taken from TanDEM-X \citep{rizzoli2017}.} 
	\label{fig:map:grouped}
\end{figure}

\subsection{Sample case study}\label{sec:results:case}
A sample case on unseen test data illustrates how the model from the deep learning approach \emph{sees} a specific weather event.
In the afternoon of June 20, 2019, a weak upper level trough embedded in southwesterly flow passed over the Alps, whereas below crest height the flow was predominantly around the Alps.
Lightning in the target area (Fig.~\ref{fig:mapcase}) occured in the warm sector in a zone with the highest values of equivalent potential temperature.
Its accompanying front had just arrived on the west coast of Europe.
The lightning model correctly identified lightning in the eastern half of Fig.~\ref{fig:mapcase} while misclassifying several occurences in its western half.

\begin{figure}
	\centering
	\includegraphics[width=0.8\textwidth, keepaspectratio]{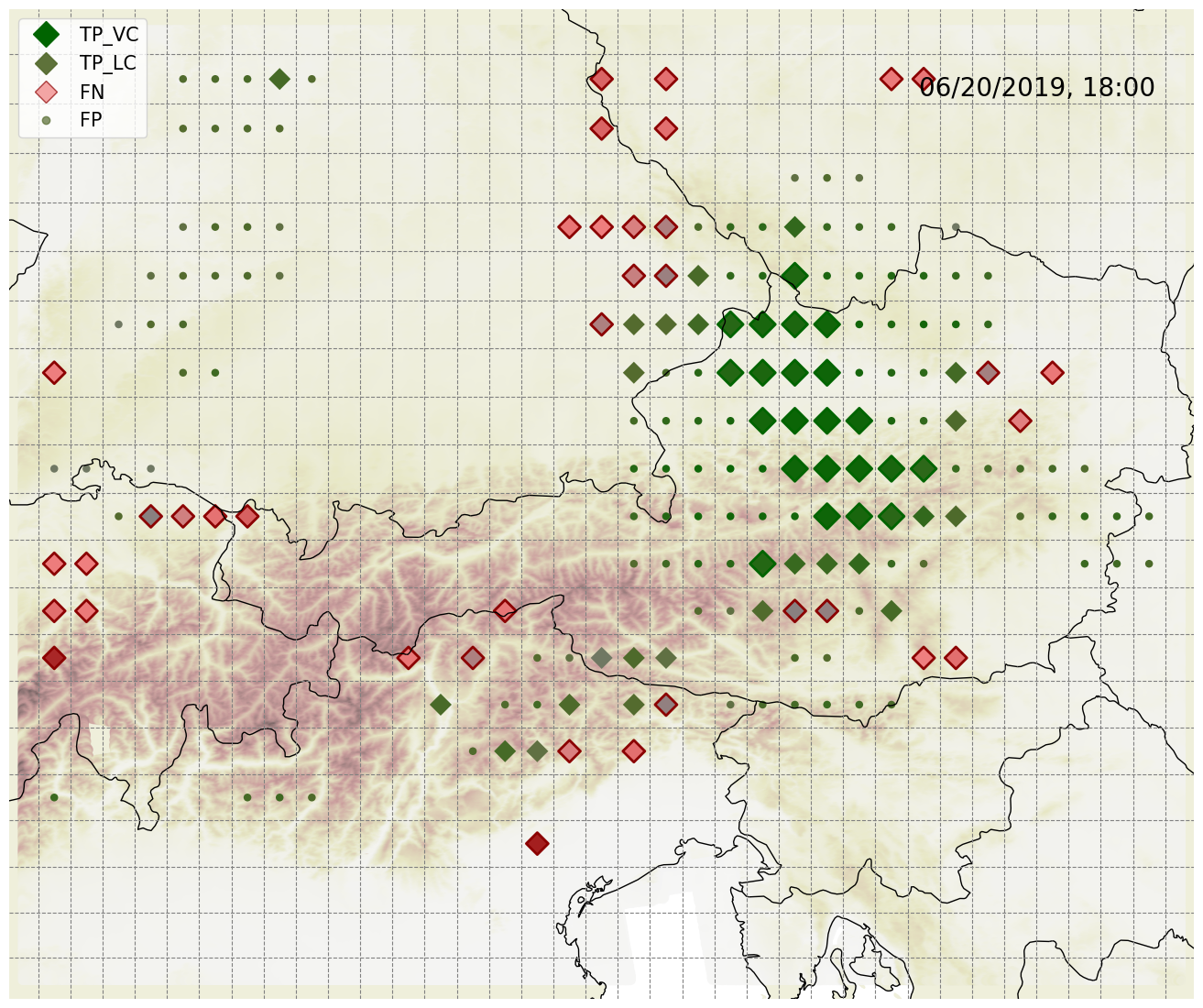}
	\caption{The map shows ERA5 grid cells with classifications of true positive (green diamonds), false negative (red diamonds) and false positive (dots)
	for the test data case June 20, 2019, in the hour before 18:00 UTC which is a case of the unseen test data. The size of the green diamonds indicates
	whether it is a \emph{very} or \emph{less} confident true positive.
	Low saturation of the red diamonds indicates
	that the output of the network was close to labeling the cell as one with lightning activity.
	The data for the displayed topography layer is taken from TanDEM-X \citep{rizzoli2017}.}
	\label{fig:mapcase}
\end{figure}

\conclusions[Discussion and Conclusions]\label{sec:discussion}  
In this study a neural network is trained on the vertical columns of raw ERA5 data without inducing any further expert knowledge about atmospheric processes to classify whether there was a lightning event or not.
Then SHAP values are used to explain which variables and vertical levels attribute the most to correct classifications of cells with lightning activity.
As indicated in Sect.~\ref{sec:results:expl}, the specific snow water and ice water content significantly capture attention, with peak interest occurring at a geopotential height of approximately $4000\,\operatorname{m}$ and $7000\,\operatorname{m}$ (cswc), and at heights of $9000\,\operatorname{m}$ and $11000\,\operatorname{m}$ (ciwc) respectively.
The neural network discovered by itself the essential ingredient for lightning, namely charge separation.
It occurs when ice cyrstals (ciwc) and larger frozen particles (graupel, cswc) are present in the convective updraft.
Once the graupel is sufficiently heavy, its velocity is smaller than the velocity of the rising ice crystals, and the collisions between ice crystals and graupel result in oppositely charged particles \citep{reynolds1957, saunders2006}.
Fig.~1 in \citet{lopez2016} shows the typical distribution of charges in a mature thunderstorm cloud.
However, it is noteworthy that the model seems to be particularly interested in the cloud ice water content at a height of $9000$ and $11000\,\operatorname{m}$ while recent literature usually looks at the cloud ice water content at $440\,\operatorname{hPa}$ (typically about $6000\,\operatorname{m}$) \citep{finney2014, finney2018, silva2022}.
Focusing on the region close to the tropopause between $9000$ and $11000\,\operatorname{m}$ means that it is crucial to vent ice particles all the way up to the tropopause and form anvils as is typical of the thunderstorm clouds.

Moreover, the model leverages the presence of southerly winds and vertical updrafts as reliable indicators for lightning occurrence especially in the northwestern Po Plain. Additionally, high specific humidity below $4000\,\operatorname{m}$ serves as a robust proxy in the central and eastern Po Plain, as well as in the southern regions of the Slovenian Alps.

The results in this work suggest promising future applications.
Being able to train a neural network directly on atmospheric soundings with good ability to distinguish between cells with and without lightning activity, and then opening the black box may enable researchers to gain a better understanding of atmospheric processes in regions like e.g.\ equatorial Africa where ample studies are scarce \citep{chakraborty2022}.
The first MGT-I satellite was launched on 13 December 2022 and will provide a lightning imager \citep{holmlund2021} which appears to be a promising source for the target variable.
Furthermore, many existing models come with two very different parameterizations for ocean and land \citep{finney2014} and this inevitably leads to discontinuities in coastal areas.
Also the reasons for the much lower lightning frequency over ocean are not as well understood yet.
Explainable AI might be a valuable building block in moving towards a more holistic understanding of the underlying atmospheric processes.

Future work might improve the results presented in this study. 
Convection and cloud processes are not purely vertical processes and thus ML parameterization greatly benefits from using multiple neighboring vertical atmospheric columns instead of a single column.
\citet{wang2022} work with $192\,\operatorname{km} \times 192\,\operatorname{km}$ grid cells to model, among others, subgrid zonal and meridional momentum flux due to vertical advection and suggest that a $3 \times 3$ subgrid could further improve the performance of the deep learning approach.
Here, a simple fully connected neural network is used and therefore the model loses information about the connectivity of the values along the levels of the vertical profiles.
Using convolutional layers to process the profiles would, most likely, further improve the results. 
However, the goal of this work was to use a very simple machine learning approach to detect cells with lightning activity and then to disect the model to understand which atmospheric conditions the model has found to be typical for lightning. 
The input data was preprocessed with only very little meteorological expertise to ensure that the methodology is easily transferable to other regions of the earth where the understanding of lightning related atmospheric processes is still scarce.

\codedataavailability{The software (version 1.1; Python and R code) used to produce the results and plots in this manuscript is licenced under MIT and published on GitHub \url{https://github.com/noxthot/xai_lightningprocesses} \citep{sw:xailightning}.
The source code relies on two data sources:
\begin{enumerate}
  \item ERA5 \citep{hersbach2020} data are available via the Climate Data Store \citep{data:era5_sfc, data:era5_ml}.
Scripts for sending the retrievals are included in the \texttt{data-preprocessing} directory of the GitHub repository \citep{sw:xailightning}.
  \item The ALDIS data \citep{schulz2016}, which are the second important source of data,
cannot be made available to the public. However, ALDIS data are
available on request from ALDIS aldis@ove.at -- fees may be charged.
\end{enumerate}}

\authorcontribution{\textbf{Gregor Ehrensperger}: Methodology, Software - model \& explainable AI \& plotting \& data preparation, Writing – original draft. \textbf{Thorsten Simon}: Data curation, Software - reference model \& plotting, Writing – original draft. \textbf{Georg Mayr}: Supervision, Writing - review \& editing. \textbf{Tobias Hell}: Conceptualization, Methodology.}

\competinginterests{The authors declare that they have no known competing financial
interests or personal relationships that could have appeared to influence
the work reported in this paper.}

\begin{acknowledgements}
We are grateful for data support by Gerhard Diendorfer and Wolfgang Schulz from OVE-ALDIS.
We thank Deborah Morgenstern and Johannes Horak for their script to compute
geopotential height on ERA5 model levels. 
Also, we thank Johanna Rissbacher for contributing parts of Fig.~\ref{fig:mapcase} and the corresponding code.
\end{acknowledgements}

\financialsupport{This work was funded by the Austrian Science Fund (FWF, grant no.~P\,31836) and the Austrian Research Promotion Agency (FFG, grant no.~872656).}

\bibliographystyle{copernicus}
\bibliography{references}

\begin{thebibliography}{60}
\providecommand{\natexlab}[1]{#1}
\providecommand{\url}[1]{\texttt{#1}}
\providecommand{\urlprefix}{}
\expandafter\ifx\csname urlstyle\endcsname\relax
  \providecommand{\doi}[1]{https://doi.org/\discretionary{}{}{}#1}\else
  \providecommand{\doi}{https://doi.org/\discretionary{}{}{}\begingroup
  \urlstyle{rm}\Url}\fi

\bibitem[{Aas et~al.(2021)Aas, Jullum, and L{\o}land}]{aas2021}
Aas, K., Jullum, M., and L{\o}land, A.: Explaining Individual Predictions when
  Features are Dependent: {M}ore Accurate Approximations to {Shapley} Values,
  Artificial Intelligence, 298, 103\,502, \doi{10.1016/j.artint.2021.103502},
  2021.

\bibitem[{Allen and Pickering(2002)}]{allen2002}
Allen, D.~J. and Pickering, K.~E.: Evaluation of Lightning Flash Rate
  Parameterizations for Use in a Global Chemical Transport Model, Journal of
  Geophysical Research: Atmospheres, 107, ACH 15--1--ACH 15--21,
  \doi{10.1029/2002JD002066}, 2002.

\bibitem[{Barnes et~al.(2020)Barnes, Toms, Hurrell, Ebert-Uphoff, Anderson, and
  Anderson}]{barnes2020}
Barnes, E.~A., Toms, B., Hurrell, J.~W., Ebert-Uphoff, I., Anderson, C., and
  Anderson, D.: Indicator Patterns of Forced Change Learned by an Artificial
  Neural Network, Journal of Advances in Modeling Earth Systems, 12,
  e2020MS002\,195, \doi{10.1029/2020MS002195}, 2020.

\bibitem[{Becerra et~al.(2018)Becerra, Long, Schulz, and
  Thottappillil}]{becerra2018}
Becerra, M., Long, M., Schulz, W., and Thottappillil, R.: On the Estimation of
  the Lightning Incidence to Offshore Wind Farms, Electric Power Systems
  Research, 157, 211--226, \doi{10.1016/j.epsr.2017.12.008}, 2018.

\bibitem[{Brisson et~al.(2021)Brisson, Blahak, Lucas-Picher, Purr, and
  Ahrens}]{brisson2021}
Brisson, E., Blahak, U., Lucas-Picher, P., Purr, C., and Ahrens, B.:
  Contrasting Lightning Projection Using the Lightning Potential Index Adapted
  in a Convection-Permitting Regional Climate Model, Climate Dynamics, 57,
  2037--2051, \doi{10.1007/s00382-021-05791-z}, 2021.

\bibitem[{Brook et~al.(1982)Brook, Nakano, Krehbiel, and Takeuti}]{brook1982}
Brook, M., Nakano, M., Krehbiel, P., and Takeuti, T.: The electrical structure
  of the hokuriku winter thunderstorms, Journal of Geophysical Research:
  Oceans, 87, 1207--1215, \doi{10.1029/JC087iC02p01207}, 1982.

\bibitem[{Cecil et~al.(2014)Cecil, Buechler, and Blakeslee}]{cecil2014}
Cecil, D.~J., Buechler, D.~E., and Blakeslee, R.~J.: Gridded Lightning
  Climatology from {TRMM-LIS} and {OTD}: {D}ataset Description, Atmospheric
  Research, 135, 404--414, \doi{10.1016/j.atmosres.2012.06.028}, 2014.

\bibitem[{Chakraborty et~al.(2022)Chakraborty, Menghal, Harshitha, and
  Sodunke}]{chakraborty2022}
Chakraborty, R., Menghal, P., Harshitha, M., and Sodunke, M.: Climatology of
  Lightning Activities Across the {E}quatorial {A}frican Region, in: 2022 3rd
  URSI Atlantic and Asia Pacific Radio Science Meeting (AT-AP-RASC), pp. 1--4,
  IEEE, \doi{10.23919/AT-AP-RASC54737.2022.9814276}, 2022.

\bibitem[{Charn and Parishani(2021)}]{charn2021}
Charn, A.~B. and Parishani, H.: Predictive Proxies of Present and Future
  Lightning in a Superparameterized Model, Journal of Geophysical Research:
  Atmospheres, 126, \doi{10.1029/2021JD035461}, 2021.

\bibitem[{Cummins et~al.(1998)Cummins, Krider, and Malone}]{cummins1998}
Cummins, K., Krider, E., and Malone, M.: The {US} {N}ational {L}ightning
  {D}etection {N}etwork and Applications of Cloud-to-Ground Lightning Data by
  Electric Power Utilities, IEEE Transactions on Electromagnetic Compatibility,
  40, 465--480, \doi{10.1109/15.736207}, 1998.

\bibitem[{DeCaria et~al.(2005)DeCaria, Pickering, Stenchikov, and
  Ott}]{decaria2005}
DeCaria, A.~J., Pickering, K.~E., Stenchikov, G.~L., and Ott, L.~E.:
  Lightning-Generated $\mathrm{NO}_x$ and its Impact on Tropospheric Ozone
  Production: A Three-Dimensional Modeling Study of a
  {S}tratosphere-{T}roposphere {E}xperiment: {R}adiation, {A}erosols and
  {O}zone ({STERAO-A}) Thunderstorm, Journal of Geophysical Research:
  Atmospheres, 110, \doi{10.1029/2004JD005556}, 2005.

\bibitem[{Dutta and Pal(2022)}]{dutta2022}
Dutta, D. and Pal, S.~K.: Interpretation of Black Box for Short-Term
  Predictions of Pre-Monsoon Cumulonimbus Cloud Events over {K}olkata, Journal
  of Data, Information and Management, 4, 167--183,
  \doi{10.1007/s42488-022-00071-9}, 2022.

\bibitem[{Ehrensperger et~al.(2024)Ehrensperger, Hell, Mayr, and
  Simon}]{sw:xailightning}
Ehrensperger, G., Hell, T., Mayr, G., and Simon, T.: xai\_lightningprocesses,
  \doi{10.5281/zenodo.10899180}, 2024.

\bibitem[{Feldmann et~al.(2021)Feldmann, Germann, Gabella, and
  Berne}]{feldmann2021}
Feldmann, M., Germann, U., Gabella, M., and Berne, A.: A Characterisation of
  {A}lpine Mesocyclone Occurrence, Weather and Climate Dynamics, 2, 1225--1244,
  \doi{10.5194/wcd-2-1225-2021}, 2021.

\bibitem[{Finney et~al.(2014)Finney, Doherty, Wild, Huntrieser, Pumphrey, and
  Blyth}]{finney2014}
Finney, D.~L., Doherty, R.~M., Wild, O., Huntrieser, H., Pumphrey, H.~C., and
  Blyth, A.~M.: Using Cloud Ice Flux to Parametrise Large-Scale Lightning,
  Atmospheric Chemistry and Physics, 14, 12\,665--12\,682,
  \doi{10.5194/acp-14-12665-2014}, 2014.

\bibitem[{Finney et~al.(2018)Finney, Doherty, Wild, Stevenson, MacKenzie, and
  Blyth}]{finney2018}
Finney, D.~L., Doherty, R.~M., Wild, O., Stevenson, D.~S., MacKenzie, I.~A.,
  and Blyth, A.~M.: A Projected Decrease in Lightning under Climate Change,
  Nature Climate Change, 8, 210--213, \doi{10.1038/s41558-018-0072-6}, 2018.

\bibitem[{Groenemeijer et~al.(2019)Groenemeijer, P{\'u}cik, Tsonevsky, and
  Bechtold}]{groenemeijer2019ecmwfmemo}
Groenemeijer, P., P{\'u}cik, T., Tsonevsky, I., and Bechtold, P.: An Overview
  of Convective Available Potential Energy and Convective Inhibition provided
  by {NWP} models for operational forecasting, \doi{10.21957/q392hofrl}, 2019.

\bibitem[{Hersbach et~al.(2017)Hersbach, Bell, Berrisford, Hirahara, Horányi,
  Muñoz‐Sabater, Nicolas, Peubey, Radu, Schepers, Simmons, Soci, Abdalla,
  Abellan, Balsamo, Bechtold, Biavati, Bidlot, Bonavita, De~Chiara, Dahlgren,
  Dee, Diamantakis, Dragani, Flemming, Forbes, Fuentes, Geer, Haimberger,
  Healy, Hogan, Hólm, Janisková, Keeley, Laloyaux, Lopez, Lupu, Radnoti,
  de~Rosnay, Rozum, Vamborg, Villaume, and Thépaut}]{data:era5_ml}
Hersbach, H., Bell, B., Berrisford, P., Hirahara, S., Horányi, A.,
  Muñoz‐Sabater, J., Nicolas, J., Peubey, C., Radu, R., Schepers, D.,
  Simmons, A., Soci, C., Abdalla, S., Abellan, X., Balsamo, G., Bechtold, P.,
  Biavati, G., Bidlot, J., Bonavita, M., De~Chiara, G., Dahlgren, P., Dee, D.,
  Diamantakis, M., Dragani, R., Flemming, J., Forbes, R., Fuentes, M., Geer,
  A., Haimberger, L., Healy, S., Hogan, R., Hólm, E., Janisková, M., Keeley,
  S., Laloyaux, P., Lopez, P., Lupu, C., Radnoti, G., de~Rosnay, P., Rozum, I.,
  Vamborg, F., Villaume, S., and Thépaut, J.-N.: Complete ERA5 from 1979:
  Fifth generation of ECMWF atmospheric reanalyses of the global climate,
  \urlprefix\url{https://cds.climate.copernicus.eu/#!/home}, accessed on
  27-05-2021, 2017.

\bibitem[{Hersbach et~al.(2018)Hersbach, Bell, Berrisford, Biavati, Horányi,
  Muñoz~Sabater, Nicolas, Peubey, Radu, Rozum, Schepers, Simmons, Soci, Dee,
  and Thépaut}]{data:era5_sfc}
Hersbach, H., Bell, B., Berrisford, P., Biavati, G., Horányi, A.,
  Muñoz~Sabater, J., Nicolas, J., Peubey, C., Radu, R., Rozum, I., Schepers,
  D., Simmons, A., Soci, C., Dee, D., and Thépaut, J.-N.: ERA5 hourly data on
  single levels from 1959 to present, \doi{10.24381/cds.adbb2d47}, accessed on
  16-02-2022, 2018.

\bibitem[{Hersbach et~al.(2020)Hersbach, Bell, Berrisford, Hirahara, Horányi,
  Muñoz-Sabater, Nicolas, Peubey, Radu, Schepers, Simmons, Soci, Abdalla,
  Abellan, Balsamo, Bechtold, Biavati, Bidlot, Bonavita, De~Chiara, Dahlgren,
  Dee, Diamantakis, Dragani, Flemming, Forbes, Fuentes, Geer, Haimberger,
  Healy, Hogan, Hólm, Janisková, Keeley, Laloyaux, Lopez, Lupu, Radnoti,
  de~Rosnay, Rozum, Vamborg, Villaume, and Thépaut}]{hersbach2020}
Hersbach, H., Bell, B., Berrisford, P., Hirahara, S., Horányi, A.,
  Muñoz-Sabater, J., Nicolas, J., Peubey, C., Radu, R., Schepers, D., Simmons,
  A., Soci, C., Abdalla, S., Abellan, X., Balsamo, G., Bechtold, P., Biavati,
  G., Bidlot, J., Bonavita, M., De~Chiara, G., Dahlgren, P., Dee, D.,
  Diamantakis, M., Dragani, R., Flemming, J., Forbes, R., Fuentes, M., Geer,
  A., Haimberger, L., Healy, S., Hogan, R.~J., Hólm, E., Janisková, M.,
  Keeley, S., Laloyaux, P., Lopez, P., Lupu, C., Radnoti, G., de~Rosnay, P.,
  Rozum, I., Vamborg, F., Villaume, S., and Thépaut, J.-N.: The {ERA5} Global
  Reanalysis, Quarterly Journal of the Royal Meteorological Society, 146,
  1999--2049, \doi{10.1002/qj.3803}, 2020.

\bibitem[{Hilburn et~al.(2021)Hilburn, Ebert-Uphoff, and Miller}]{hilburn2021}
Hilburn, K.~A., Ebert-Uphoff, I., and Miller, S.~D.: Development and
  Interpretation of a Neural-Network-Based Synthetic Radar Reflectivity
  Estimator Using {GOES-R} Satellite Observations, Journal of Applied
  Meteorology and Climatology, 60, 3--21, \doi{10.1175/JAMC-D-20-0084.1}, 2021.

\bibitem[{Holle(2016)}]{holle2016}
Holle, R.~L.: A Summary of Recent National-Scale Lightning Fatality Studies,
  Weather, Climate, and Society, 8, 35--42, \doi{10.1175/WCAS-D-15-0032.1},
  2016.

\bibitem[{Holmlund et~al.(2021)Holmlund, Grandell, Schmetz, Stuhlmann, Bojkov,
  Munro, Lekouara, Coppens, Viticchie, August, Theodore, Watts, Dobber, Fowler,
  Bojinski, Schmid, Salonen, Tjemkes, Aminou, and Blythe}]{holmlund2021}
Holmlund, K., Grandell, J., Schmetz, J., Stuhlmann, R., Bojkov, B., Munro, R.,
  Lekouara, M., Coppens, D., Viticchie, B., August, T., Theodore, B., Watts,
  P., Dobber, M., Fowler, G., Bojinski, S., Schmid, A., Salonen, K., Tjemkes,
  S., Aminou, D., and Blythe, P.: {M}eteosat {T}hird {G}eneration ({MTG}):
  {C}ontinuation and Innovation of Observations from Geostationary Orbit,
  Bulletin of the American Meteorological Society, 102, 990--1015,
  \doi{10.1175/BAMS-D-19-0304.1}, 2021.

\bibitem[{Houze(2012)}]{houze2012}
Houze, R.~A.: Orographic Effects on Precipitating Clouds, Reviews of
  Geophysics, 50, 1--47, \doi{10.1029/2011RG000365}, 2012.

\bibitem[{Lopez(2016)}]{lopez2016}
Lopez, P.: A Lightning Parameterization for the {ECMWF} Integrated Forecasting
  System, Monthly Weather Review, 144, 3057--3075,
  \doi{10.1175/MWR-D-16-0026.1}, 2016.

\bibitem[{Lou et~al.(2012)Lou, Caruana, and Gehrke}]{lou2017}
Lou, Y., Caruana, R., and Gehrke, J.: Intelligible Models for Classification
  and Regression, in: Proceedings of the 18th ACM SIGKDD International
  Conference on Knowledge Discovery and Data Mining, KDD '12, pp. 150--158,
  Association for Computing Machinery, New York, NY, USA, ISBN 9781450314626,
  \doi{10.1145/2339530.2339556}, 2012.

\bibitem[{Lundberg and Lee(2017)}]{lundberg2017}
Lundberg, S.~M. and Lee, S.-I.: A Unified Approach to Interpreting Model
  Predictions, in: Advances in Neural Information Processing Systems, edited by
  Guyon, I., Luxburg, U.~V., Bengio, S., Wallach, H., Fergus, R., Vishwanathan,
  S., and Garnett, R., vol.~30, Curran Associates, Inc.,
  \urlprefix\url{https://proceedings.neurips.cc/paper/2017/file/8a20a8621978632d76c43dfd28b67767-Paper.pdf},
  2017.

\bibitem[{Mayer and Barnes(2021)}]{mayer2021}
Mayer, K.~J. and Barnes, E.~A.: Subseasonal Forecasts of Opportunity Identified
  by an Explainable Neural Network, Geophysical Research Letters, 48,
  e2020GL092\,092, \doi{10.1029/2020GL092092}, 2021.

\bibitem[{McCaul et~al.(2009)McCaul, Goodman, LaCasse, and Cecil}]{mccaul2009}
McCaul, E.~W., Goodman, S.~J., LaCasse, K.~M., and Cecil, D.~J.: Forecasting
  Lightning Threat Using Cloud-Resolving Model Simulations, Weather and
  Forecasting, 24, 709--729, \doi{10.1175/2008WAF2222152.1}, 2009.

\bibitem[{Morgenstern et~al.(2022)Morgenstern, Stucke, Simon, Mayr, and
  Zeileis}]{morgenstern2022}
Morgenstern, D., Stucke, I., Simon, T., Mayr, G.~J., and Zeileis, A.:
  Differentiating Lightning in Winter and Summer with Characteristics of the
  Wind Field and Mass Field, Weather and Climate Dynamics, 3, 361--375,
  \doi{10.5194/wcd-3-361-2022}, 2022.

\bibitem[{Morgenstern et~al.(2023)Morgenstern, Stucke, Mayr, Zeileis, and
  Simon}]{morgenstern2023}
Morgenstern, D., Stucke, I., Mayr, G.~J., Zeileis, A., and Simon, T.:
  Thunderstorm environments in Europe, Weather and Climate Dynamics, 4,
  489--509, \doi{10.5194/wcd-4-489-2023}, 2023.

\bibitem[{Murray(2018)}]{murray2018}
Murray, L.~T.: An Uncertain Future for Lightning, Nature Climate Change, 8,
  191--192, \doi{10.1038/s41558-018-0094-0}, 2018.

\bibitem[{Price and Rind(1992)}]{price1992}
Price, C. and Rind, D.: A Simple Lightning Parameterization for Calculating
  Global Lightning Distributions, Journal of Geophysical Research: Atmospheres,
  97, 9919--9933, \doi{10.1029/92JD00719}, 1992.

\bibitem[{Reineking et~al.(2010)Reineking, Weibel, Conedera, and
  Bugmann}]{reineking2010}
Reineking, B., Weibel, P., Conedera, M., and Bugmann, H.: Environmental
  Determinants of Lightning- v.\ Human-Induced Forest Fire Ignitions Differ in
  a Temperate Mountain Region of {S}witzerland, International Journal of
  Wildland Fire, 19, 541--557, \doi{10.1071/WF08206}, 2010.

\bibitem[{Reynolds et~al.(1957)Reynolds, Brook, and Gourley}]{reynolds1957}
Reynolds, S., Brook, M., and Gourley, M.~F.: Thunderstorm Charge Separation,
  Journal of Atmospheric Sciences, 14, 426--436,
  \doi{10.1175/1520-0469(1957)014<0426:TCS>2.0.CO;2}, 1957.

\bibitem[{Ritenour et~al.(2008)Ritenour, Morton, McManus, Barillo, and
  Cancio}]{ritenour2008}
Ritenour, A.~E., Morton, M.~J., McManus, J.~G., Barillo, D.~J., and Cancio,
  L.~C.: Lightning Injury: {A} Review, Burns, 34, 585--594,
  \doi{10.1016/j.burns.2007.11.006}, 2008.

\bibitem[{Rizzoli et~al.(2017)Rizzoli, Martone, Gonzalez, Wecklich, {Borla
  Tridon}, Bräutigam, Bachmann, Schulze, Fritz, Huber, Wessel, Krieger, Zink,
  and Moreira}]{rizzoli2017}
Rizzoli, P., Martone, M., Gonzalez, C., Wecklich, C., {Borla Tridon}, D.,
  Bräutigam, B., Bachmann, M., Schulze, D., Fritz, T., Huber, M., Wessel, B.,
  Krieger, G., Zink, M., and Moreira, A.: Generation and performance assessment
  of the global TanDEM-X digital elevation model, ISPRS Journal of
  Photogrammetry and Remote Sensing, 132, 119--139,
  \doi{https://doi.org/10.1016/j.isprsjprs.2017.08.008}, 2017.

\bibitem[{Romps et~al.(2018)Romps, Charn, Holzworth, Lawrence, Molinari, and
  Vollaro}]{romps2018}
Romps, D.~M., Charn, A.~B., Holzworth, R.~H., Lawrence, W.~E., Molinari, J.,
  and Vollaro, D.: {CAPE} Times {P} Explains Lightning Over Land But Not the
  Land-Ocean Contrast, Geophysical Research Letters, 45, 12,623--12,630,
  \doi{10.1029/2018GL080267}, 2018.

\bibitem[{Saunders et~al.(2006)Saunders, Bax-norman, Emersic, Avila, and
  Castellano}]{saunders2006}
Saunders, C. P.~R., Bax-norman, H., Emersic, C., Avila, E.~E., and Castellano,
  N.~E.: Laboratory Studies of the Effect of Cloud Conditions on
  Graupel/Crystal Charge Transfer in Thunderstorm Electrification, Quarterly
  Journal of the Royal Meteorological Society, 132, 2653--2673,
  \doi{10.1256/qj.05.218}, 2006.

\bibitem[{Schulz et~al.(2016)Schulz, Diendorfer, Pedeboy, and
  Poelman}]{schulz2016}
Schulz, W., Diendorfer, G., Pedeboy, S., and Poelman, D.~R.: The {E}uropean
  Lightning Location System {EUCLID} Part~1: {P}erformance Analysis and
  Validation, Natural Hazards and Earth System Sciences, 16, 595--605,
  \doi{10.5194/nhess-16-595-2016}, 2016.

\bibitem[{Shan et~al.(2023)Shan, Allen, Li, Pickering, and Lapierre}]{shan2023}
Shan, S., Allen, D., Li, Z., Pickering, K., and Lapierre, J.:
  Machine-learning-based investigation of the variables affecting summertime
  lightning occurrence over the Southern Great Plains, 23, 14\,547--14\,560,
  \doi{10.5194/acp-23-14547-2023}, 2023.

\bibitem[{Shapley(1952)}]{shapley1952}
Shapley, L.~S.: A Value for N-Person Games, RAND Corporation, Santa Monica, CA,
  \doi{10.7249/P0295}, 1952.

\bibitem[{Shi et~al.(2022)Shi, Zhang, Fan, Chen, Liu, Li, and Liu}]{shi2022}
Shi, M., Zhang, W., Fan, P., Chen, Q., Liu, Z., Li, Q., and Liu, X.: Modelling
  Deep Convective Activity Using Lightning Clusters and Machine Learning,
  International Journal of Climatology, 42, 952--973, \doi{10.1002/joc.7282},
  2022.

\bibitem[{Silva et~al.(2022)Silva, Keller, and Hardin}]{silva2022}
Silva, S.~J., Keller, C.~A., and Hardin, J.: Using an Explainable Machine
  Learning Approach to Characterize Earth System Model Errors: Application of
  {SHAP} Analysis to Modeling Lightning Flash Occurrence, Journal of Advances
  in Modeling Earth Systems, 14, e2021MS002\,881, \doi{10.1029/2021MS002881},
  2022.

\bibitem[{Simon et~al.(2023)Simon, Mayr, Morgenstern, Umlauf, and
  Zeileis}]{simon2023}
Simon, T., Mayr, G., Morgenstern, D., Umlauf, N., and Zeileis, A.:
  Amplification of annual and diurnal cycles of alpine lightning, Climate
  Dynamics, 61, 1--13, \doi{10.1007/s00382-023-06786-8}, 2023.

\bibitem[{Srivastava et~al.(2014)Srivastava, Hinton, Krizhevsky, Sutskever, and
  Salakhutdinov}]{srivastava2014}
Srivastava, N., Hinton, G., Krizhevsky, A., Sutskever, I., and Salakhutdinov,
  R.: Dropout: A Simple Way to Prevent Neural Networks from Overfitting,
  Journal of Machine Learning Research, 15, 1929--1958,
  \urlprefix\url{http://jmlr.org/papers/v15/srivastava14a.html}, 2014.

\bibitem[{Stirnberg et~al.(2021)Stirnberg, Cermak, Kotthaus, Haeffelin,
  Andersen, Fuchs, Kim, Petit, and Favez}]{stirnberg2021}
Stirnberg, R., Cermak, J., Kotthaus, S., Haeffelin, M., Andersen, H., Fuchs,
  J., Kim, M., Petit, J.-E., and Favez, O.: Meteorology-Driven Variability of
  Air Pollution ({PM~1}) Revealed with Explainable Machine Learning,
  Atmospheric Chemistry and Physics, 21, 3919--3948,
  \doi{10.5194/acp-21-3919-2021}, 2021.

\bibitem[{Takahashi et~al.(2019)Takahashi, Sugimoto, Kawano, and
  Suzuki}]{takashi2019}
Takahashi, T., Sugimoto, S., Kawano, T., and Suzuki, K.: Microphysical
  {Structure} and {Lightning} {Initiation} in {Hokuriku} {Winter} {Clouds},
  124, 13\,156--13\,181, \doi{10.1029/2018JD030227}, 2019.

\bibitem[{Takeuti et~al.(1978)Takeuti, Nakano, Brook, Raymond, and
  Krehbiel}]{takeuti1978}
Takeuti, T., Nakano, M., Brook, M., Raymond, D.~J., and Krehbiel, P.: The
  anomalous winter thunderstorms of the {Hokuriku} {Coast}, 83, 2385--2394,
  \doi{10.1029/JC083iC05p02385}, 1978.

\bibitem[{Tippett et~al.(2019)Tippett, Lepore, Koshak, Chronis, and
  Vant-Hull}]{tippett2019}
Tippett, M.~K., Lepore, C., Koshak, W.~J., Chronis, T., and Vant-Hull, B.:
  Performance of a Simple Reanalysis Proxy for {U.S.} Cloud-to-Ground
  Lightning, International Journal of Climatology, 39, 3932--3946,
  \doi{10.1002/joc.6049}, 2019.

\bibitem[{Toms et~al.(2021)Toms, Barnes, and Hurrell}]{toms2021}
Toms, B.~A., Barnes, E.~A., and Hurrell, J.~W.: Assessing Decadal
  Predictability in an Earth-System Model Using Explainable Neural Networks,
  Geophysical Research Letters, 48, e2021GL093\,842,
  \doi{10.1029/2021GL093842}, 2021.

\bibitem[{Tost et~al.(2007)Tost, J\"ockel, and Lelieveld}]{tost2007}
Tost, H., J\"ockel, P., and Lelieveld, J.: Lightning and Convection
  Parameterisations -- Uncertainties in Global Modelling, Atmospheric Chemistry
  and Physics, 7, 4553--4568, \doi{10.5194/acp-7-4553-2007}, 2007.

\bibitem[{Ukkonen and M\"akel\"a(2019)}]{ukkonen2019}
Ukkonen, P. and M\"akel\"a, A.: Evaluation of Machine Learning Classifiers for
  Predicting Deep Convection, J.\ Adv.\ Model.\ Earth\ Sy., 11, 1784--1802,
  \doi{10.1029/2018MS001561}, 2019.

\bibitem[{Ukkonen et~al.(2017)Ukkonen, Manzato, and M\"akel\"a}]{ukkonen2017}
Ukkonen, P., Manzato, A., and M\"akel\"a, A.: Evaluation of Thunderstorm
  Predictors for {F}inland Using Reanalyses and Neural Networks, ournal of
  Applied Meteorology and Climatology, 56, 2335--2352,
  \doi{10.1175/JAMC-D-16-0361.1}, 2017.

\bibitem[{{van der Velden} et~al.(2022){van der Velden}, Kuijf, Gilhuijs, and
  Viergever}]{vandervelden2022}
{van der Velden}, B.~H., Kuijf, H.~J., Gilhuijs, K.~G., and Viergever, M.~A.:
  Explainable artificial intelligence (XAI) in deep learning-based medical
  image analysis, Medical Image Analysis, 79, 102\,470,
  \doi{10.1016/j.media.2022.102470}, 2022.

\bibitem[{Wang et~al.(2021)Wang, Zheng, Wu, and Takagi}]{wang2021}
Wang, D., Zheng, D., Wu, T., and Takagi, N.: Winter {Positive}
  {Cloud}-to-{Ground} {Lightning} {Flashes} {Observed} by {LMA} in {Japan}, 16,
  402--411, \doi{10.1002/tee.23310}, 2021.

\bibitem[{Wang et~al.(2022)Wang, Yuval, and O’Gorman}]{wang2022}
Wang, P., Yuval, J., and O’Gorman, P.~A.: Non-local parameterization of
  atmospheric subgrid processes with neural networks, Journal of Advances in
  Modeling Earth Systems, p. e2022MS002984, \doi{10.1029/2022MS002984}, 2022.

\bibitem[{Wood(2017)}]{wood2017generalized}
Wood, S.~N.: {G}eneralized Additive Models: {A}n Introduction with \textsf{R},
  Texts in Statistical Science, Chapman \& Hall/CRC, Boca Raton, 2nd edn.,
  \doi{10.1201/9781420010404}, 2017.

\bibitem[{Wood et~al.(2017)Wood, Li, Shaddick, and Augustin}]{wood2017gigadata}
Wood, S.~N., Li, Z., Shaddick, G., and Augustin, N.~H.: Generalized Additive
  Models for Gigadata: {M}odeling the {U.K.\ }Black Smoke Network Daily Data,
  Journal of the American Statistical Association, 112, 1199--1210,
  \doi{10.1080/01621459.2016.1195744}, 2017.

\bibitem[{Zhang et~al.(2021)Zhang, Tiňo, Leonardis, and Tang}]{zhang2021}
Zhang, Y., Tiňo, P., Leonardis, A., and Tang, K.: A Survey on Neural Network
  Interpretability, IEEE Transactions on Emerging Topics in Computational
  Intelligence, 5, 726--742, \doi{10.1109/TETCI.2021.3100641}, 2021.

\end{thebibliography}

\end{nolinenumbers}
\end{document}